\documentclass[12pt]{iopart}

\usepackage{graphicx}

\begin{document}

\title{What is the probability of connecting two points ?}

\author{Christian Tanguy}

\address{France Telecom Division R\&D CORE, 38--40 rue du
G\'{e}n\'{e}ral Leclerc, 92794 Issy-les-Moulineaux Cedex, France}
\ead{christian.tanguy@orange-ftgroup.com}
\begin{abstract}
The two-terminal reliability, known as the pair connectedness or connectivity function in percolation theory,
may actually be expressed as a product of transfer matrices in which the probability of operation of each
link and site is exactly taken into account. When link and site probabilities are $p$ and $\rho$, it obeys an
asymptotic power-law behaviour, for which the scaling factor is the transfer matrix's eigenvalue of largest
modulus. The location of the complex zeros of the two-terminal reliability polynomial exhibits structural
transitions as $0 \leq \rho \leq 1$.
\end{abstract}


\submitto{\JPA}
\maketitle

\section{Introduction}

Since the original work of Moore and Shannon \cite{MooreShannon56}, network reliability has been a field
devoted to the calculation of the connection probability between different sites of a network constituted by
edges (links, bonds) and nodes (vertices, sites), each of them having a probability of operating correctly
(the reliability). This field, although mainly developed in an applied background \cite{Singh77}, is strongly
related to graph theory \cite{Wu82b,Biggs}, combinatorics and algebraic structures \cite{Colbourn87,Shier91},
percolation theory \cite{Grimmett99,Hughes96}, as well as numerous lattice models in statistical physics
\cite{ChangShrockA,ChangShrockB,Salas,Welsh00}. For instance, the all-terminal reliability ${\rm Rel}_A$,
i.e., the probability that all nodes are connected, is derived from the Tutte polynomial, an invariant of the
associated graph, when all edges have the same reliability $p$ ($0 \leq p \leq 1$). This polynomial appears
in the partition function for various Potts models, and has been calculated for several families of graphs
\cite{ChangShrockA,ChangShrockB,Salas}; the location of its complex zeros has also been studied
\cite{ChangShrockB,Salas,Royle04}. The two-terminal reliability ${\rm Rel}_2(s \rightarrow t)$, the
probability that a source $s$ and a destination $t$ are connected, is known in percolation theory as the
connectivity function or pair connectedness. It has been used in modeling epidemics or fire propagation
\cite{Grimmett99,Hughes96}. This approach is complementary to the effort recently devoted on complex
networks, in which the network resilience, i.e., its robustness against link or node failures (sometimes
following deliberate attacks) has been studied for `scale-free' random graphs \cite{Albert02}.

Exact reliability calculations are known to be very difficult \cite{Oxley02}, except for series-parallel
reducible graphs for which only successive simplifications \{$p_{\rm series} = p_1 \, p_2$, $p_{\rm parallel}
= p_1/\!/p_2 = p_1 + p_2 - p_1 \, p_2$\} are needed. Even for planar graphs with identical edge reliabilities
$p$ and perfect nodes (i.e., $p_{\rm node} \equiv 1$), their algorithmic complexity has been classified as
\#P-hard \cite{Colbourn87,Welsh93}. Yet, the development of Internet traffic makes it important to assess the
overall reliability of network connections, when links {\em and} nodes may fail.

\begin{figure}
\centering
\includegraphics[scale=0.5]{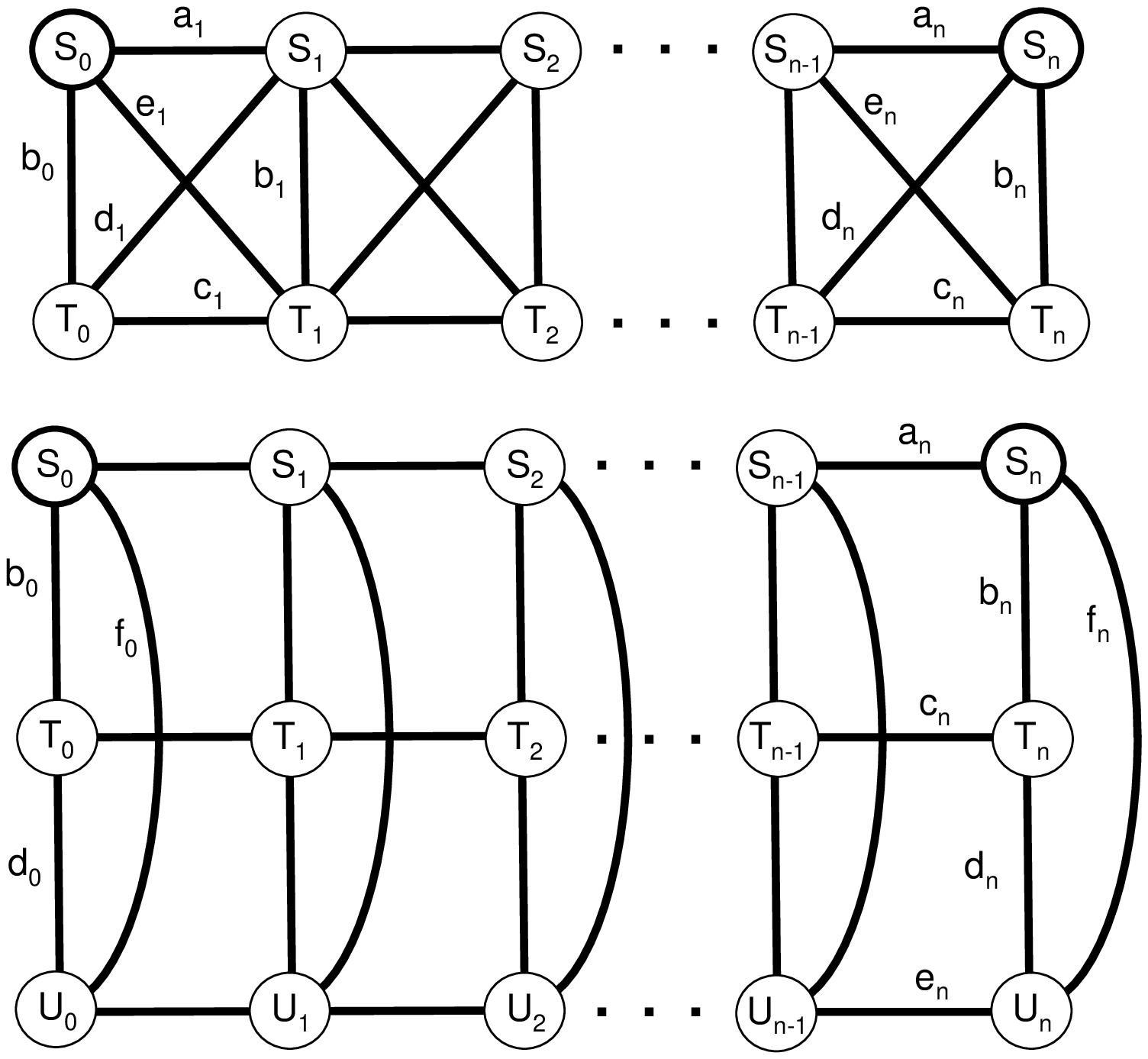}
\caption{Generic network architectures: (top) $K_4$-ladder (bottom)
$K_3$-cylinder. Links and nodes are identified by their
reliabilities: $a_n$, $b_n$, etc. for links, $S_n$, $T_n$, and $U_n$
for nodes. The source is $S_0$, the possible destinations are $S_n$,
$T_n$, or $U_n$. A missing link or node's reliability is simply set
to zero.} \label{reseaux}
\end{figure}

In this work, we show that for a network represented by an undirected graph $G$, the two-terminal reliability
may be expressed as a product of transfer matrices, where {\em individual} edge and node reliabilities are
exactly taken into account. Such a factorization, already observed for graph colouring polynomials
\cite{Biggs,Salas}, 2D-percolation in square strips \cite{Derrida80} or all-terminal reliability polynomials
\cite{ChangShrockA,ChangShrockB}, originates with the underlying algebraic structure of the graph. We apply
our method to the two examples ($K_n$ is the complete graph with $n$ nodes) of figure~\ref{reseaux}. The
$K_4$-ladder describes a generic architecture for long-haul connections, while the $K_3$-cylinder slightly
generalizes the `sponge model' of width three by Seymour and Welsh \cite{Seymour78}. When edge and node
reliabilities are respectively equal to $p$ and $\rho$, a unique transfer matrix is involved; its largest
eigenvalue determines the asymptotic power-law behaviour of reliability as a function of the ladder length.
The location of the complex zeros of ${\rm Rel}_2(p)$ exhibits striking structure transitions as $\rho$
decreases from one to zero. We illustrate the variety of behaviours for the above-mentioned graphs. For the
sake of completeness, we finally give the matrix decomposition for the all-terminal reliability of the
$K_4$-ladder with arbitrary edge reliabilities (the uniform case has already been treated by Chang and Shrock
\cite{ChangShrockA}).

\section{Graph decomposition}

The gist of our method is to simplify the graph by removing links of
the $n$th (last) elementary cell of the network, namely the
edges and nodes indexed by $n$, a procedure called pivotal
decomposition or deletion-contraction \cite{Colbourn87}. If the end
terminal $t$ (which can be regarded as perfect) is connected to node
$u$ through edge $e$, with respective reliabilities $p_u$ and $p_e$,
then
\begin{eqnarray}
\label{pivotaldecomposition} {\rm Rel}_2(G) & = & (1 - p_e) \, {\rm
Rel}_2(G \! \setminus \! e) +
p_e \, p_u \, {\rm Rel}_2(G \cdot e) \nonumber \\
& & + p_e \, (1-p_u) \, {\rm Rel}_2(G \! \setminus \! u) ,
\end{eqnarray}
where $G \! \setminus \! e$ and $G \! \setminus \! u$ are the graphs where $e$ or $u$ have been deleted, and $G \cdot e$
the graph where $t$ and $u$ have been merged through the `contraction' of $e$;
(\ref{pivotaldecomposition}) merely sums probabilities of disjoint events. This procedure, along with
standard series-parallel reductions, is repeated for the three (instead of the usual two) secondary graphs in
order to take advantage of a structural recursivity of the graph. After a finite number of such reductions,
we get replicas of the original graph, albeit with one less elementary cell and with the $(n-1)$th
cell's edge and node reliabilities possibly renormalized by those of the $n$th cell, or set to either
zero or one. In order to ensure the existence of a recursion relation, the graph structure must be {\em
closed} under successive applications of (\ref{pivotaldecomposition}); it may initially require the use
of extra edges with symbolic reliabilities, so that all nodes of an elementary cell are connected pair-wise,
even if such links do not exist in the graph under consideration. At this point, a recursion hypothesis is
needed, giving for instance ${\rm Rel}_2(S_0 \! \rightarrow \! S_n)$ as a sum over specific polynomials in
the reliabilities indexed by $n$; these are often obvious from the $n=2$ value. Going from $n-1$ to $n$
provides the transfer matrix linking the prefactors of the polynomials, because ${\rm Rel}_2$ is an affine
function of each component reliability; the (often trivial) $n=1$ case serves as the initial condition of the
recurrence.

\section{Application to the $K_4$-ladder}
\label{K4ladder}


\begin{figure}
\centering
\includegraphics[scale=0.5]{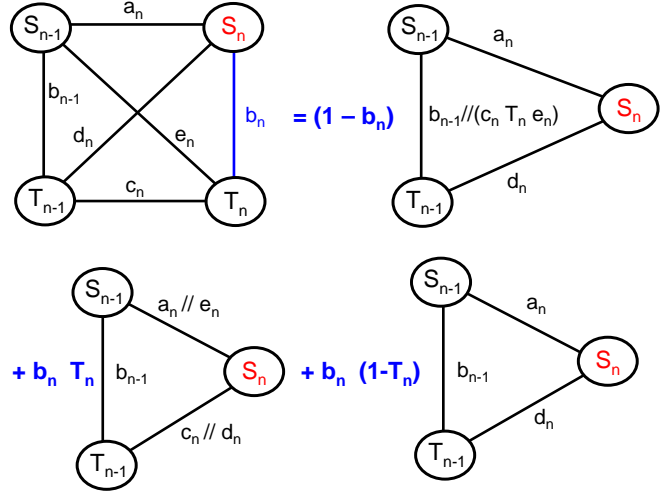}
\caption{First step of the pivotal decomposition : the removal of edge $b_n$. Three structurally identical, secondary graphs are obtained.}
\label{PivotFirstStep}
\end{figure}

Let us first illustrate this method by calculating ${\mathcal R}_n = {\rm Rel}_2(S_0 \! \rightarrow \! S_n)$
for the $K_4$-ladder (top of figure~\ref{reseaux}). Following the guidelines of the preceding section, we first
consider $b_n$ for deletion as detailed in figure~\ref{PivotFirstStep}. Note that the three secondary graphs have essentially the same structure. The renewed application of (\ref{pivotaldecomposition}) leads to two families of contributions. The
first one is a sum of ${\mathcal R}_{n-1}$-like terms with prefactors, in which the `old' $a_{n-1}$, ...,
$T_{n-1}$ are renormalized by one or more of the `new' $a_{n}$, ..., $T_{n}$. The second one is a sum of
${\rm Rel}_2(S_0 \! \rightarrow \! T_{n-1})$-like terms. There is no need for coupled recursion relations for
the two destinations $S_n$ and $T_n$, since they are essentially identical through the permutations $a_n \!
\leftrightarrow \! e_n$, $c_n \! \leftrightarrow \! d_n$, and $S_n \! \leftrightarrow \! T_n$. ${\mathcal
R}_n$ may be expressed as the sum of five polynomials in $a_n$, ..., $T_n$ (see below). The five prefactors
at step $n$ are obtained from those at step $n-1$ by a recursion relation which translates as a $5 \times 5$
transfer matrix (such calculations are routinely performed by mathematical software). The value of ${\mathcal
R}_{1}$ leads to
\begin{eqnarray}
\label{Rel2S0Snformelle} {\mathcal R}_n & = & (1 \; 0 \; 0 \; 0 \;
0) \, M_n \, M_{n-1} \, \cdots M_1 \, M_0 \, \left(
\begin{array}{c}
1 \\
0 \\
0 \\
0 \\
0 \end{array} \right) ,
\end{eqnarray}
where $M_i$'s coefficients $M_{k l}$ are ($\overline{x} \equiv 1 - x$)
\numparts
\begin{eqnarray}
M_{1 1} & = & ( a_{i} + b_{i} \, e_{i} \, T_{i} - a_{i} \, b_{i} \, e_{i} \, T_{i}) \, S_{i} , \label{ElementsMatriceEchelleK4debut} \\
M_{1 2} & = & ( d_{i} + b_{i} \, c_{i} \, T_{i} - d_{i} \, b_{i} \, c_{i} \, T_{i}) \, S_{i} , \\
M_{1 3} & = & a_i \, d_i \, S_i + b_i \, (\chi_i + c_i \, e_i) \, S_i \, T_i , \\
M_{1 4} & = & - M_{4 4} = a_{i} \, \, e_{i} \, M_{4 2} , \\
M_{1 5} & = & - M_{4 5} = c_{i} \, \, d_{i} \, M_{4 1} , \\
M_{2 1} & = & ( e_{i} + b_{i} \, a_{i} \, S_{i} - e_{i} \, b_{i} \, a_{i} \, S_{i}) \, T_{i} , \\
M_{2 2} & = & ( c_{i} + b_{i} \, d_{i} \, S_{i} - c_{i} \, b_{i} \, d_{i} \, S_{i}) \, T_{i} , \\
M_{2 3} & = & c_i \, e_i \, T_i + b_i \, (\chi_i + a_i \, d_i) \, S_i \, T_i , \\
M_{2 4} & = & - M_{5 4} = a_{i} \, \, e_{i} \, M_{5 2} , \\
M_{2 5} & = & - M_{5 5} = c_{i} \, \, d_{i} \, M_{5 1} , \\
M_{3 1} & = & - (a_{i} \, b_{i} + a_{i} \, e_{i} + b_{i} \, e_{i} - 2 \, a_{i} \, b_{i} \, e_{i}) \, S_{i} \, T_{i} , \\
M_{3 2} & = & - (b_{i} \, c_{i} + b_{i} \, d_{i} + c_{i} \, d_{i} - 2 \, b_{i} \, c_{i} \, d_{i}) \, S_{i} \, T_{i} , \\
M_{3 3} & = & \left( (1-2 \, b_i) \, \chi_i - b_i \, (c_i \, e_i + a_i \, d_i)\right) \, S_i \, T_i , \\
M_{3 4} & = & - M_{1 4} - M_{2 4} , \\
M_{3 5} & = & - M_{1 5} - M_{2 5} , \\
M_{4 1} & = & \overline{a_{i}} \, \overline{b_{i}} \, e_{i} \, S_{i} \, T_{i} , \\
M_{4 2} & = & \overline{b_{i}} \, c_{i} \, \overline{d_{i}} \, S_{i} \, T_{i} , \\
M_{4 3} & = & \overline{b_{i}} \, (\chi_i + c_i \, e_i) \, S_i \, T_i , \\
M_{5 1} & = & a_{i} \, \overline{b_{i}} \, \overline{e_{i}} \, S_{i} \, T_{i} , \\
M_{5 2} & = & \overline{b_{i}} \, \overline{c_{i}} \, d_{i} \, S_{i} \, T_{i} , \\
M_{5 3} & = & \overline{b_{i}} (\chi_i + a_i \, d_i) \, S_i \, T_i , \label{ElementsMatriceEchelleK4fin}
\end{eqnarray}
\endnumparts with $ \chi_i = \overline{a_{i}} \, \overline{c_{i}} \, d_{i} \, e_{i} + a_{i} \, c_{i} \,
\overline{d_{i}} \, \overline{e_{i}} - a_{i} \, c_{i} \, d_{i} \, e_{i}$. In the $n=0$ case, $a_0 = 1$ and
$c_0=d_0=e_0=0$. The five above-mentioned polynomials are actually given by the first row of $M_n$. ${\rm
Rel}_2(S_0 \! \rightarrow \! T_n)$ is given by (\ref{Rel2S0Snformelle}) if the left vector is $(0 \; 1 \; 0
\; 0 \; 0)$. We have here another useful instance of a product of random matrices \cite{Crisanti93}.


The case of identical reliabilities $a_i = \cdots = e_i = p$ (unless $i=0$, see the restriction above) and
$S_i=T_i=\rho$ is worth investigating, since only the $n$th power of a unique matrix needs be taken.
Because of the recursion relation between successive values of ${\mathcal R}_n$, the generating function
${\mathcal G}(z) = \sum_{n=0}^{\infty} \, {\mathcal R}_n \, z^n $ is a rational fraction of $z$. Its
denominator ${\mathcal D}(z)$ is derived from the characteristic polynomial of the transfer matrix, taken at
$1/z$. The numerator of ${\mathcal G}(z)$ is then deduced from the computed first terms of the ${\mathcal G}(z) \, {\mathcal D}(z)$'s expansion. The final result reads:
\begin{eqnarray}
{\mathcal G}(z) & = & \frac{1}{2} \, \rho \, (1 - p \, \rho) +
\frac{{\mathcal N}(z)}{{\mathcal D}(z)} , \label{GechelleK4}\\
{\mathcal N}(z) & = & \frac{1}{2} \, \rho \, (1 + p \, \rho) \nonumber \\
& & - \frac{1}{2} \, p^2 \, \rho^3 \, (2 - 10 \, p +13 \, p^2-4 \,
p^3
- p^3 \, \rho) \, z \nonumber \\
& & + (1-p)^2 \, p^5 \, (2 - 4 \, p +p^2) \, (1 - \rho) \, \rho^5 \,
z^2 , \label{NechelleK4}
\\
{\mathcal D}(z) & = & 1 - p \, \rho\, \left( 2 + 4\,p\,\rho -
14\,p^2\,\rho + 13\,p^3\,\rho  - 4\,p^4\,\rho  \right) \, z \nonumber \\
& &+ 2 \, \left( 1 - p \right) \,p^3 \, {\rho }^3 \, \left( 2 - 7\,p
+ 4\,p^2 + 7\,p^2\,\rho \right. \nonumber \\
& & \left. \hskip2.5cm
- 10\,p^3\,\rho  + 5\,p^4\,\rho  - p^5\,\rho  \right) \, z^2 \nonumber \\
& &  - 4\,\left( 2 - p \right) \,{\left( 1 - p \right)
}^3\,p^6\,\left( 1 - \rho  \right) \,{\rho }^5 \, z^3 .
\label{DechelleK4}
\end{eqnarray}
Equations~(\ref{GechelleK4}--\ref{DechelleK4}) are simpler for
perfect nodes, because the denominator is of degree 2 in $z$; a
partial fraction decomposition provides
\begin{eqnarray}
{\mathcal R}_n & = & \frac{1-p}{2} \, \delta_{n,0} + a_+ \,
\lambda_+^n + a_- \, \lambda_-^n ,
\\
\lambda_{\pm} & = & \frac{p}{2} \, \left[ 2 + 4 \, p - 14 \, p^2 +13 \, p^3 - 4 \, p^4 \pm \sqrt{{\mathcal A}} \right] , \\
a_{\pm} & = &  \frac{1+p}{4} \pm \frac{2 + 2 \, p +10 \, p^2 -27 \,
p^3 +19 \, p^4-4 \, p^5}{4 \, \sqrt{{\mathcal A}} } , \nonumber \\
& & \\
 {\mathcal A} & = & 4 + 32 \, p^2 -204 \, p^3 +452 \, p^4 - 516
\,
p^5 \nonumber \\
& &  + 329 \, p^6 -112 \, p^7 +16 \, p^8 .
\end{eqnarray}
As $n$ grows, ${\mathcal R}_n \approx a_+ \, \lambda_+^n$: the two-terminal reliability exhibits a power-law
behaviour, the scaling factor being $\lambda_+$, the eigenvalue of largest modulus. Alternatively, ${\mathcal
R}_n \sim \exp(-n/\xi)$, where $\xi = -1/\ln(\lambda_+)$ is the correlation length of percolation theory
\cite{Grimmett99}.

\begin{figure}
\centering
\includegraphics[scale=1]{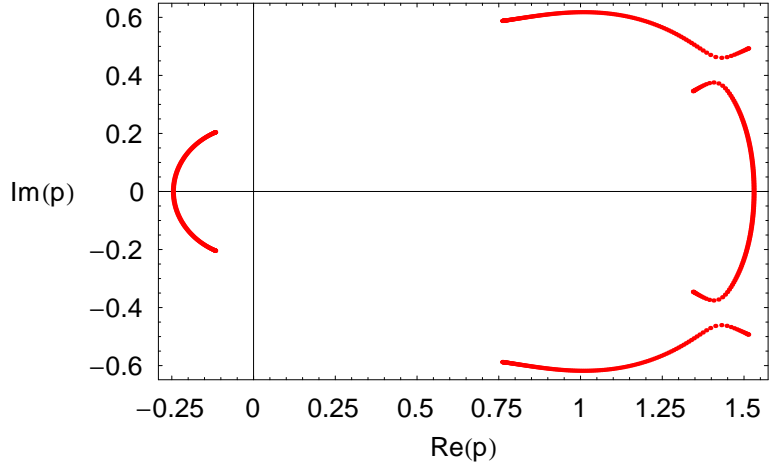} \caption{Location of the complex zeros of the
two-terminal reliability polynomial ${\mathcal R}_n(p,\rho)$ for $n=150$ and $\rho = 1$.}
\label{ZerosRel2K4rho1.0}
\end{figure}

\begin{figure}
\centering
\includegraphics[scale=1]{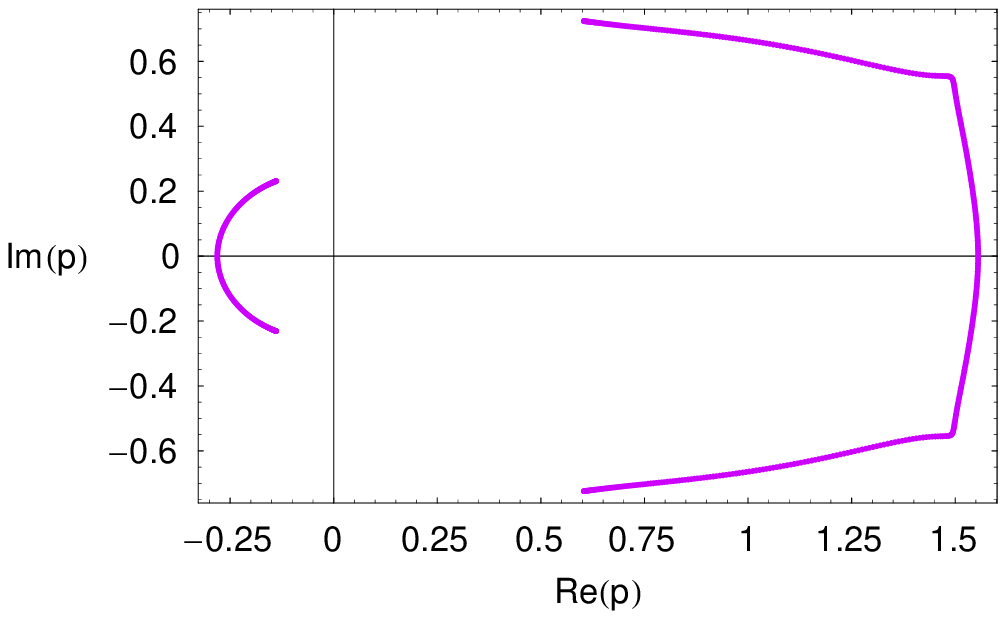} \caption{Same as
figure~\ref{ZerosRel2K4rho1.0}, with $\rho = 0.8$.}
\label{ZerosRel2K4rho0.8}
\end{figure}

\begin{figure}
\centering
\includegraphics[scale=1]{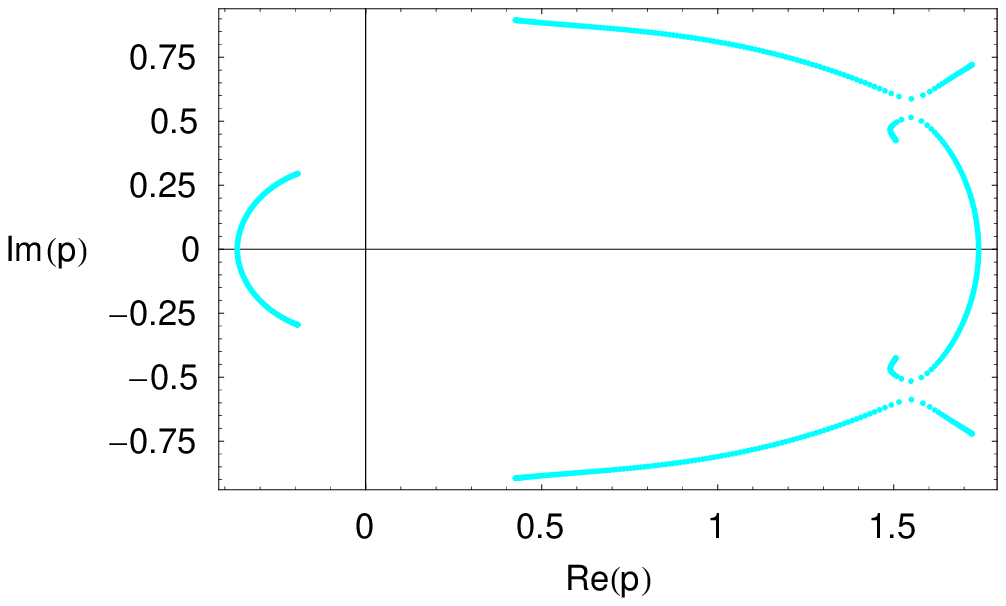} \caption{Same as
figure~\ref{ZerosRel2K4rho1.0}, with $\rho = 0.5$.}
\label{ZerosRel2K4rho0.5}
\end{figure}

\begin{figure}
\centering
\includegraphics[scale=1]{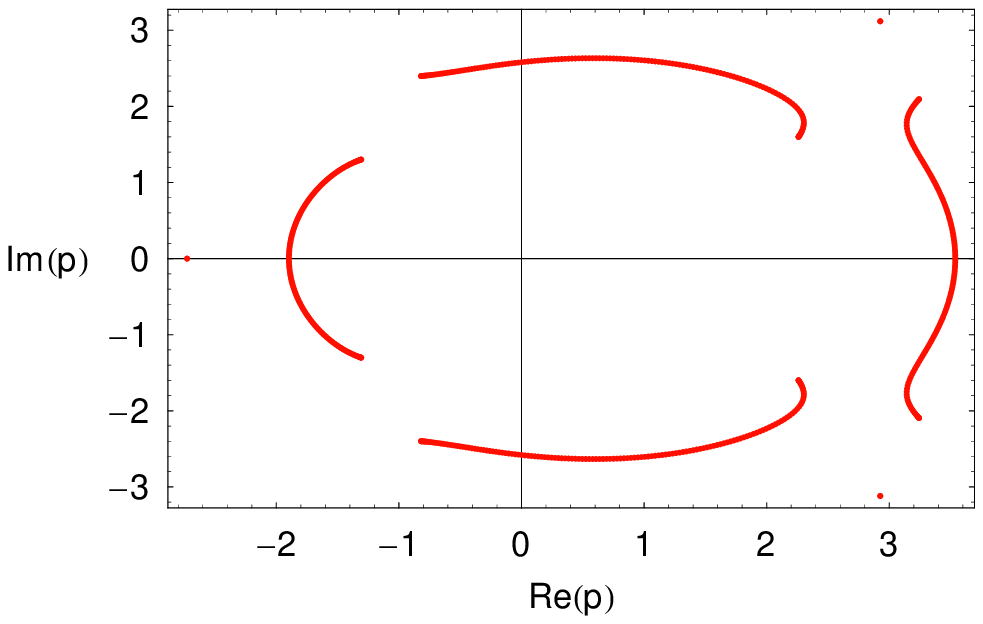}
\caption{Same as figure~\ref{ZerosRel2K4rho1.0}, with $\rho = 0.01$.}
\label{ZerosRel2K4rho0.01}
\end{figure}



The location of the zeros of ${\rm Rel}_2(p)$ in the complex plane is also worth investigating. The situation
differs from that for chromatic \cite{Biggs,Salas} and all-terminal polynomials \cite{ChangShrockA}, because
${\rm Rel}_2(p)$ is not a graph invariant. However, the node reliability $\rho$ is an extra parameter that
has a deep impact on the curves to which the zeros of ${\rm Rel}_2(p)$ converge as $n \rightarrow \infty$.
The critical values of $\rho$ at which shape transitions occur may be deduced \cite{Beraha78} from the three
roots of ${\mathcal D}(1/z)$. The straightforward but tedious procedures used to determine these values, along with a few asymptotic expansions as $\rho \to 0$, are outlined in the Appendix (they are also applied to the $K_3$-cylinder configuration). We limit ourselves to the final results in the following sections.

A sample of the richness of behaviour is displayed in
figures~\ref{ZerosRel2K4rho1.0}--\ref{ZerosRel2K4rho0.01} for the $K_4$-ladder and decreasing values of
$\rho$. We initially observe four well-separated `curves' that merge into two when $\rho$ is exactly equal
to 0.8 (see figure~\ref{ZerosRel2K4rho0.8}) and separate again. When $\rho$ further decreases, other isolated
zeros appear, as in figure~\ref{ZerosRel2K4rho0.01}. These zeros occur in pairs, the separation of which
vanishes exponentially with $n$, and converge to roots of the algebraic equation
\begin{eqnarray}
0 & = & 2 +2 \, \rho +4 \, (3 \, \rho+1 ) \, \rho \, p -(40 \, \rho+11) \, \rho \, p^2 \nonumber \\ & & +(45
\, \rho+4 ) \, \rho \, p^3-20 \, \rho^2 \, p^4 + 3 \, \rho^2 \, p^5 \, .
\label{equation points isoles}
\end{eqnarray}
Equation (\ref{equation points isoles}) is obtained by ensuring that ${\mathcal N}(z)$ and ${\mathcal D}(z)$
have a common root. The true limiting isolated points are such that this root is the eigenvalue of greatest
complex modulus at the given $p$ and $\rho$. Actually, the triplet of figure~\ref{ZerosRel2K4rho0.01} appears
only when $\rho < \rho_{{\rm c}_1} \approx 0.175221381869$, where $\rho_{{\rm c}_1}$ is a solution of (see the Appendix)
\begin{eqnarray}
0 & = & -32768 - 198656\,\rho  + 3990544\,{\rho }^2 - 12843528\,{\rho }^3 \nonumber \\
& & + 16258037\,{\rho }^4 - 6757568\,{\rho }^5 - 2015436\,{\rho }^6 - 575540\,{\rho }^7 \nonumber \\
& & + 4636356\,{\rho }^8 - 3082436\,{\rho }^9 + 624640\,{\rho }^{10} \, ,
\label{equation rho critique K4 troisieme zero isole}
\end{eqnarray}
whereas the associated $p_{{\rm c}_1} \approx -0.604692601721$ is a solution of
\begin{eqnarray}
0 & = & 40 - 364\,p + 1064\,p^2 - 700\,p^3 - 1946\,p^4 + 4296\,p^5 \nonumber \\
& & - 3465\,p^6 + 1074\,p^7 + 146\,p^8 - 176\,p^9 + 32\,p^{10} \, .
\label{equation p critique K4 troisieme zero isole}
\end{eqnarray}
If $\rho_{{\rm c}_1} < \rho < \rho_{{\rm c}_2} \approx 0.406657811123$ (the algebraic equation satisfied by
$\rho_{{\rm c}_2}$ is actually of degree 65 in $\rho$), only the two rightmost isolated points are present.

The leftmost isolated point, located on the real negative axis, is asymptotically given by $ - (2 \,
\rho)^{-1/3} + 25/24 + \Or(\rho^{1/3})$; for the other two, $\rho$ must be replaced by $\rho \, e^{\pm 2 \,
\rmi \, \pi}$. By contrast, the algebraic curves' asymptotic limit is a circle of radius $(2 \, \rho)^{-1/4}$
centred at $(27/32,0)$, demonstrating a different power-law behaviour with $\rho$. Finally, a third critical
value $\rho_{{\rm c}_3} \approx 0.49137068$ also appears, for which we have not been able to find the
defining algebraic equation satisfied by $\rho$ (its degree is likely to be large); at this value, there is
an asymptotic (anti-)crossing of the curves in the vicinity of $p \approx 1.55533445 + \rmi \, 0.55314582$.

\section{$K_3$-cylinder}
\label{K3cylinder}

In the second architecture of figure~\ref{reseaux}, $S_0$ is still the source while $S_n$, $T_n$, and $U_n$ are
the three possible destinations (the last two are equivalent through a permutation of variables). The crucial
point is to take all $f_i \neq 0$, because in the successive applications of
(\ref{pivotaldecomposition}), the merging of nodes entails a secondary graph in which $S_{n-1}$ and
$U_{n-1}$ are connected. As mentioned above, the dummy --- with respect to the Manhattan-like strip --- link
$f_n$ between $S_{n}$ and $U_{n}$ must therefore be present right at the start; this allows us to unveil the
{\em coupled} recursion relations between the source and {\em all} the destinations. Each source-destination
reliability is a sum of eight polynomials in reliabilities indexed by $n$. This could lead to $24 \times 24$
transfer matrices $\widetilde{M}_i$. However, several rows of these matrices, if not identical, are linearly dependent;  rearrangements of terms actually reduce their size to $13 \times 13$, even when $f_i=0$.

The final result reads
\begin{eqnarray}
\label{Rel2S0Snformellegenerale} {\widetilde{\mathcal R}}_n & = &  {\mathbf v}_L \, \widetilde{M}_n \,
\widetilde{M}_{n-1} \, \cdots \widetilde{M}_1 \, \widetilde{M}_0 \, {\mathbf v}_R ,
\end{eqnarray}
where ${\mathbf v_R}$ is the column vector defined by $({\mathbf v_R})_k = \delta_{k 1}$ for $1 \leq k \leq
13$ (using the Kronecker notation: $\delta_{k l}$ is equal to 1 if $k = l$ and 0 if $k \neq l$), and
${\mathbf v_L}$ is a row vector which depends on the destination: $\left({\mathbf v}_{S_n}\right)_k =
\delta_{1 k}$, $\left({\mathbf v}_{T_n}\right)_k = \delta_{2 k}$, and $\left({\mathbf v}_{U_n}\right)_k =
\delta_{3 k}$. The matrix elements are much lengthier than in
(\ref{ElementsMatriceEchelleK4debut}--\ref{ElementsMatriceEchelleK4fin}), and are given in the Appendix for
the sake of completeness.

\subsection{$f_i = 0$} \label{fn nul}

Following the procedure outlined in the preceding section, we can compute the new generating function. For
perfect nodes, $\widetilde{{\mathcal G}}(S_0 \! \rightarrow \! U_n)$ is given by $\widetilde{{\mathcal
N}}/(\widetilde{{\mathcal D}}_1 \, \widetilde{{\mathcal D}}_2)$:
\begin{eqnarray}
\widetilde{{\mathcal N}} & = & p^2 - \left( 1 - p \right) \,p^4\,
   \left( 3 + 3\,p - 4\,p^2 \right) \,z \nonumber \\
   & & \hskip-1cm +
  {\left( 1 - p \right) }^3\,p^6\,
   \left( 2 + 11\,p - 3\,p^2 - 2\,p^3 \right) \,z^2 \nonumber \\
& & \hskip-1cm + {\left( 1 - p \right) }^3\,p^8\,
   \left( 2 - 4\,p + 3\,p^2 + 11\,p^3 - 13\,p^4 +
     3\,p^5 \right) \,z^3 \nonumber  \\
& & \hskip-1cm - {\left( 1 - p \right) }^4\,p^{10}\,
   \left( 3 + 6\,p - 12\,p^2 + 10\,p^3 - 10\,p^4 +
     4\,p^5 \right) \,z^4 \nonumber  \\
& & \hskip-1cm + {\left( 1 - p \right) }^6\,p^{12}\,
   \left( 1 + 8\,p - p^2 - 5\,p^3 - p^4 + p^5 \right)
     \,z^5 \nonumber  \\
& & \hskip-1cm - {\left( 1 - p \right) }^8\,p^{15}\,
   \left( 2 + 5\,p - 4\,p^2 \right) \,z^6 +
  {\left( 1 - p \right) }^{10}\,p^{18}\,z^7 , \\
\widetilde{{\mathcal D}}_1 & = &  1 - \left( 1 - p^2 \right) \,p \,
   \left( 1 + p - p^2 \right) \,z \nonumber  \\
   & & \hskip-1cm +
  {\left( 1 - p \right) }^2\,p^3\,
   \left( 1 + p + p^2 - 2\,p^3 \right) \,z^2
   - {\left( 1 - p \right) }^4\,p^6\,z^3 ,
\\
\widetilde{{\mathcal D}}_2 & = & 1 - p\,\left( 2 + 2\,p + p^2 -
9\,p^3 + 5\,p^4 \right)
     \,z \nonumber  \\
& &  + \left( 1 - p \right) \,p^2\,
   \left( 1 + 5\,p + 5\,p^2 - 6\,p^3 - 15\,p^4 \right. \nonumber \\
   & & \left. \hskip2.2cm +
     13\,p^5 + p^6 - 2\,p^7 \right) \,z^2 \nonumber  \\
& &  - {\left( 1 - p \right) }^2\,p^4\,
   \left( 2 + 6\,p + 6\,p^2 - 26\,p^3 + 17\,p^4 \right. \nonumber \\
   & & \left. \hskip2.2cm -
     18\,p^5 + 27\,p^6 - 16\,p^7 + 3\,p^8 \right) \,z^3 \nonumber  \\
& &  + {\left( 1 - p \right) }^4\,p^6\,
   \left( 1 + 6\,p + 4\,p^2 - p^3 - 17\,p^4 \right. \nonumber \\
   & & \left. \hskip2.2cm +
     9\,p^5 + 3\,p^6 - 2\,p^7 \right) \,z^4 \nonumber  \\
& &  -
  {\left( 1 - p \right) }^6\,p^9\,
   \left( 2 + 4\,p + p^2 - 7\,p^3 + 3\,p^4 \right) \,
   z^5 \nonumber  \\
& &  + {\left( 1 - p \right) }^8\,p^{12}\,z^6 .
\end{eqnarray}
When $\rho \neq 1$, the degrees of $\widetilde{{\mathcal N}}$, $\widetilde{{\mathcal D}}_1$,
$\widetilde{{\mathcal D}}_2$ are still 7, 3 and 6, respectively; their expressions are only lengthier.

The eigenvalue of greatest modulus $\lambda_{\rm max}$ involved in the asymptotic power-law behaviour obeys
$\widetilde{{\mathcal D}}_2(1/\lambda_{\rm max}) = 0$. The degree of the denominator leads us to expect that
the `width' of the network should drastically affect the size of the transfer matrices.

\begin{figure}
\centering
\includegraphics[scale=1]{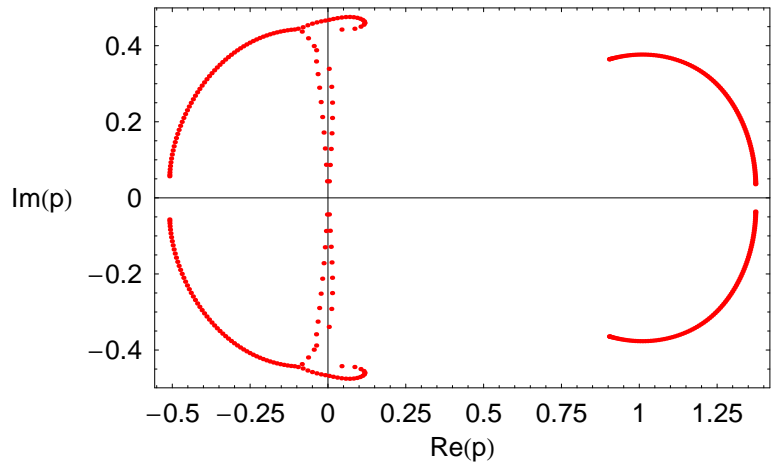}
\caption{Complex zeros of the two-terminal reliability polynomial for the $K_3$-cylinder, $f_i = 0$,
$n=100$ and $\rho = 1$.}
\label{ZerosStreet3NfnEQ0rho1.0}
\end{figure}

\begin{figure}
\centering
\includegraphics[scale=1]{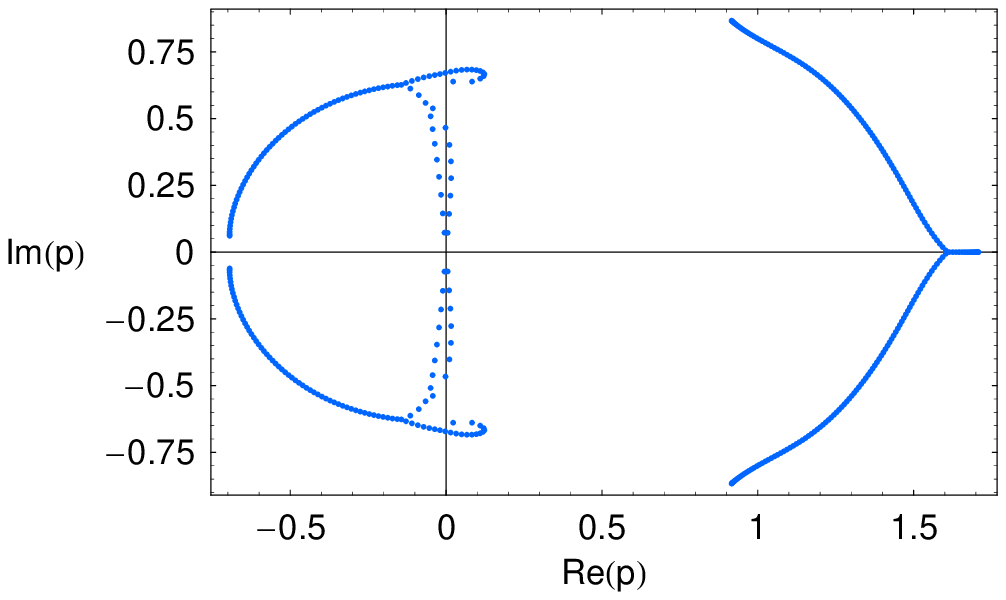}
\caption{Same as figure~\ref{ZerosStreet3NfnEQ0rho1.0}, with $\rho = 0.6$.}
\label{ZerosStreet3NfnEQ0rho0.6}
\end{figure}

\begin{figure}
\centering
\includegraphics[scale=1]{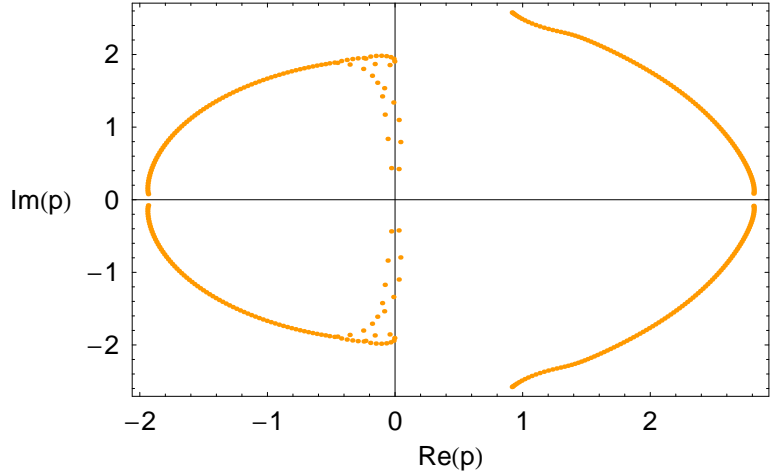}
\caption{Same as figure~\ref{ZerosStreet3NfnEQ0rho1.0}, with $\rho = 0.1$.}
\label{ZerosStreet3NfnEQ0rho0.1}
\end{figure}

The associated complex zeros are displayed for various values of $\rho$ in
figures~\ref{ZerosStreet3NfnEQ0rho1.0}--\ref{ZerosStreet3NfnEQ0rho0.1}. The overall structure is more
complicated than that for the $K_4$-ladder, but some features are quite similar.

A segment of the real axis appears as a limit curve when $0.4202958 < \rho < 0.8092264$. These critical
values obey different criteria. Indeed, the higher one (with the associated critical, real $p \approx
1.53039659$) occurs when two complex roots of $\widetilde{{\mathcal D}}_2(z)$ have the same (lowest) modulus
as a real negative root of $\widetilde{{\mathcal D}}_1(z)$. By contrast, the lower critical value 0.4202958
appears when $\widetilde{{\mathcal D}}_2(z)$ exhibits two complex roots and a real positive root with the
same modulus (the critical $p$ is about 1.8363587).

What happens when $\rho \to 0$ ? The outermost parts of the curves tend asymptotically to a circle of radius
$\displaystyle \frac{\left(5 - \sqrt{17} \right)^{1/4}}{\sqrt{2 \, \rho}}$, i.e., approximately
$\displaystyle \frac{0.684261}{\sqrt{\rho}}$. The closed curve on the left survives. For instance, a triple
point $p_t$ goes asymptotically as $\displaystyle \pm \frac{\rmi \, a}{\sqrt{\rho}} + b$, with $a \approx
0.4610389$ and $b \approx -0.08457522$ ($a^2$ is a root of a polynomial of degree 10, and $b$ is a rational
fraction of $a$).
From each of these points, two curves head back to the origin. One of them crosses the imaginary axis at
$\displaystyle p \sim \pm \frac{\rmi \, a'}{\sqrt{\rho}}$, with $a' \approx 0.33529987$ (${a'}^2$ is actually
the root of a polynomial of degree 17).

\subsection{$f_i \neq 0$} \label{fn non nul}

In this case, the generating function $\widetilde{{\mathcal G}'}(S_0 \! \rightarrow \! U_n)$ is now equal to
$\widetilde{{\mathcal N}'}/(\widetilde{{\mathcal D}'}_1 \, \widetilde{{\mathcal D}'}_2)$, where for perfect
nodes
\begin{eqnarray}
\widetilde{{\mathcal N}'} & = & p\,( 1 + p - p^2 ) \nonumber  \\
& &  -
  ( 2 - p ) \,{( 1 - p ) }^2\,
   p^3\,( 1 + p ) \,
   ( 1 + 3\,p - 3\,p^2 ) \,z \nonumber  \\
& & +
  {( 1 - p ) }^5\,p^5\,
   ( 1 + 10\,p + 8\,p^2 - 5\,p^3 - 2\,p^4 )
     \,z^2 \nonumber  \\
& & - {( 1 - p ) }^6\,p^8\,
   ( 3 + 8\,p - 25\,p^2 + 9\,p^3 + 4\,p^4 -
     p^5 ) \,z^3 \nonumber  \\
& & +
  {( 1 - p ) }^8\,p^{11}\,
   ( 1 - 2\,p ) \,
   ( 3 + 3\,p - 7\,p^2 + 2\,p^3 ) \,z^4 \nonumber \\
& & -
  {( 1 - p ) }^{11}\,p^{14}\,
   ( 1 - 3\,p + p^2 ) \,z^5
 , \\
\widetilde{{\mathcal D}'}_1 & = &  1 - {\left( 1 - p \right) }^2\,p\,
   \left( 1 + p \right) \,
   \left( 1 + p - p^2 \right) \,z \nonumber  \\
& & +
  {\left( 1 - p \right) }^4\,p^3\,
   \left( 1 + p + p^2 - 2\,p^3 \right) \,z^2 \nonumber  \\
& & -
  {\left( 1 - p \right) }^7\,p^6\,z^3 ,
\\
\widetilde{{\mathcal D}'}_2 & = & 1 - p\,( 1 + 3\,p + 4\,p^2 - 23\,p^3 + 23\,p^4 -
     7\,p^5 ) \,z \nonumber  \\
& & +
  {( 1 - p ) }^2\,p^3\,
   ( 1 + 6\,p + 2\,p^2 - 9\,p^3 \nonumber  \\
& & \hskip2.2cm - 8\,p^4 +
     16\,p^5 - 6\,p^6 ) \,z^2 \nonumber  \\
& &  -
  {( 1 - p ) }^4\,p^6\,
   ( 2 + 4\,p + p^2 - 15\,p^3 + 12\,p^4 -
     3\,p^5 ) \,z^3 \nonumber  \\
& &  +
  {( 1 - p ) }^7\,p^9\,z^4 .
\end{eqnarray}

\begin{figure}
\centering
\includegraphics[scale=1]{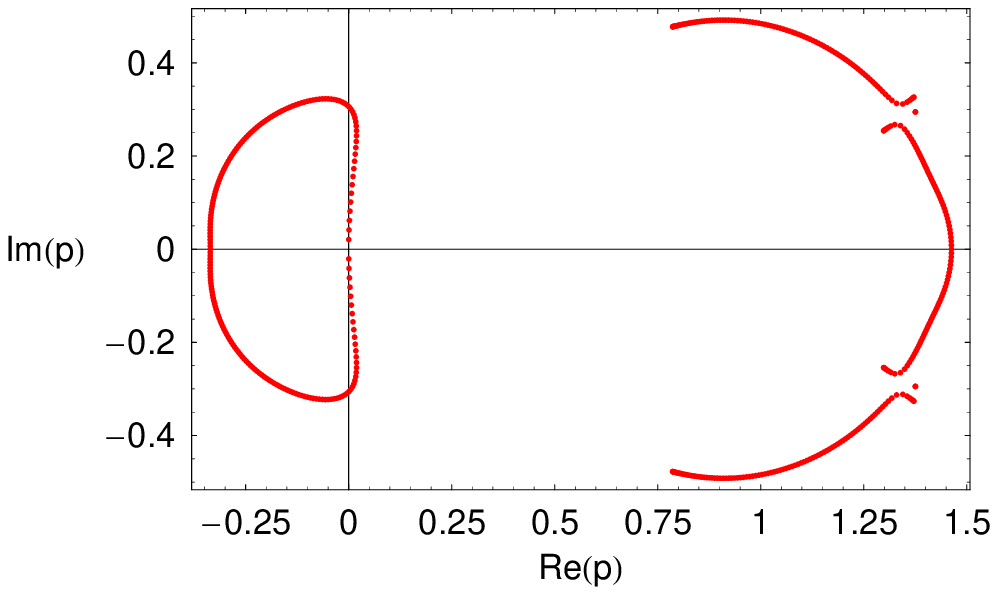}
\caption{Complex zeros of the two-terminal reliability polynomial for the $K_3$-cylinder, $f_i \neq 0$,
$n=100$ and $\rho = 1$.}
\label{ZerosStreet3NfnNEQ0rho1.0}
\end{figure}

\begin{figure}
\centering
\includegraphics[scale=1]{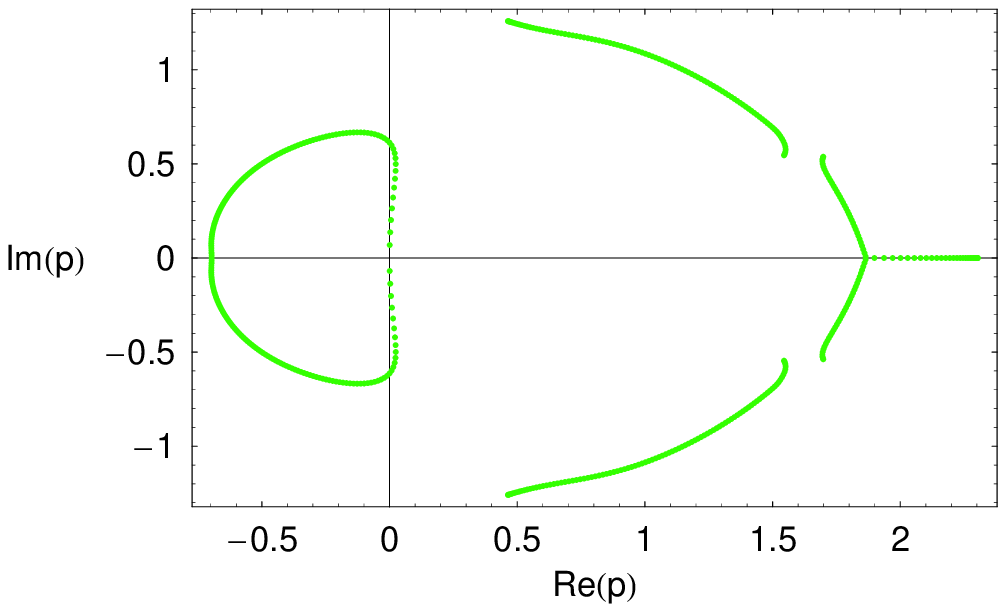}
\caption{Same as figure~\ref{ZerosStreet3NfnNEQ0rho1.0}, with $\rho = 0.3$.}
\label{ZerosStreet3NfnNEQ0rho0.3}
\end{figure}

\begin{figure}
\centering
\includegraphics[scale=1]{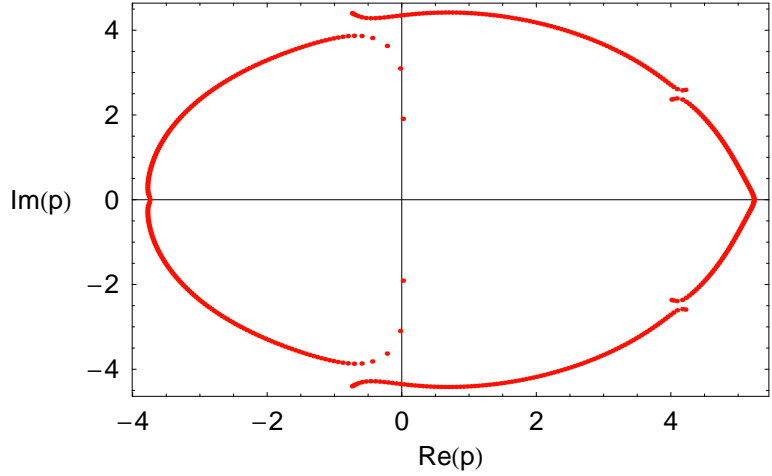}
\caption{Same as figure~\ref{ZerosStreet3NfnNEQ0rho1.0}, with $\rho = 0.01$.}
\label{ZerosStreet3NfnNEQ0rho0.01}
\end{figure}
Note that $\widetilde{{\mathcal D}'}_2$ is of degree 4 in $z$ (even when $\rho \neq 1$), so that a complete
analytical solution for the two-terminal reliability could be obtained --- but would be very cumbersome.

The location of complex zeros are displayed in
figures~\ref{ZerosStreet3NfnNEQ0rho1.0}--\ref{ZerosStreet3NfnNEQ0rho0.01}. Critical values of different
nature occur in this case. Two isolated zeros exist as long as $0.3633122889 < \rho \leq 1$. They do not
survive in the $\rho \to 0$ limit, in contrast with the $K_4$ case. For $\rho \approx 0.3633122889$, they
merge with a continuous curve at $p \approx 1.466816 \pm \rmi \, 0.5823927$. For these isolated points, the
relevant $p$ and $\rho$ obey the polynomial constraint
\begin{eqnarray}
0 & = & -2\,\left( 1 - 2\,p \right) \,
   \left( 1 + 5\,p - 4\,p^2 \right)  \nonumber \\
& & -
  p\,\left( 1 + 39\,p - 172\,p^2 + 316\,p^3 -
     230\,p^4 + 56\,p^5 \right) \,\rho  \nonumber \\
& & +
  p^2\,\left( 12 + 42\,p - 416\,p^2 + 947\,p^3 -
     899\,p^4 \right. \nonumber \\
& & \left. \hskip1cm + 382\,p^5 - 60\,p^6 \right) \,{\rho }^2
   \nonumber \\
& & + p^3\,\left( 27 - 45\,p + 33\,p^2 - 90\,p^3 +
     135\,p^4 \right. \nonumber \\
& & \left. \hskip1cm - 77\,p^5 + 15\,p^6 \right) \,{\rho }^3 ,
\label{equation points isoles K3 fi neq 0}
\end{eqnarray}
the origin of which is similar to that of (\ref{equation points isoles}).

Another feature is the segment on the real axis (see figure~\ref{ZerosStreet3NfnNEQ0rho0.3}) which occurs
when $0.016301418 < \rho < 0.83140245$. These two critical values are actually solutions of a polynomial in
$\rho$ of degree 95, and the associated critical $p$'s, namely 1.60638989 and 4.56013168, are also roots of a
polynomial in $p$ of degree 95. These transitions occur when the equation $\widetilde{{\mathcal D}'}_2(z) =
0$ has a double, real (negative) root, the opposite of which is also a root.

As in the preceding subsection, the global structure expands as $\rho \to 0$. The outer curves tend
asymptotically to a circle of radius $\displaystyle \frac{1}{7^{1/5} \, \rho^{2/5}} \approx 0.677611 \,
\rho^{-2/5}$. The closed curve on the left also survives (see figure \ref{ZerosStreet3NfnNEQ0rho0.01}). Here
again, it crosses the imaginary axis asymptotically at $\displaystyle p \sim \pm \rmi \, \left(
\frac{10^{-1/6}}{\rho^{1/3}} - \frac{3}{5 \, \sqrt{10}}\right)$.

\section{Transfer matrices for the all-terminal reliability ${\rm Rel}_A$}
\label{K4ladderRelA}

Nodes may be viewed as perfect in this case since the node reliabilities can be factored out, and simpler
calculations may be done because (\ref{pivotaldecomposition}) has one less term. For the $K_4$-ladder,
the transfer matrix is $2 \times 2$:
\begin{equation}
{\rm Rel}_A(n) = (1 \, 0) \; \widehat{M}_{n} \; \cdots \;
\widehat{M}_{n-1} \; \widehat{M}_{0} \left(
\begin{array}{c}
1 \\
0
\end{array}
\right) ;
\end{equation}
The matrix elements $(\widehat{M}_i)_{k l}$ of $\widehat{M}_{i}$ are ($\overline{x} \equiv 1 - x$)
\begin{eqnarray}
(\widehat{M}_i)_{1 1} & = & [(a_i+e_i) \, (c_i+d_i) - 2 \, a_i \, c_i \, d_i \, e_i] \, \overline{b_i} \nonumber \\
& & + [(a_i/\!/e_i) + (c_i/\!/d_i)] \, b_i , \\
(\widehat{M}_i)_{1 2} & = & a_i \, c_i \, d_i \, e_i \, [\frac{1}{a_i} + \frac{1}{c_i} + \frac{1}{d_i} + \frac{1}{e_i} -3] \, \overline{b_i} \nonumber \\
& & + [(a_i/\!/e_i) \, (c_i/\!/d_i)] \, b_i , \\
(\widehat{M}_i)_{2 1} & = &  [(a_i/\!/c_i) + (d_i/\!/e_i) - 2 \, (a_i \, e_i/\!/c_i \, d_i)] \, \overline{b_i} \nonumber \\
& & - (\widehat{M}_i)_{1 1} , \\
(\widehat{M}_i)_{2 2} & = &  (c_i+d_i - 2 \, c_i \, d_i) \, (a_i+e_i - 2 \, a_i \, e_i)  \, \overline{b_i} \nonumber \\
& & - (\widehat{M}_i)_{1 2} ;
\end{eqnarray}
in $\widehat{M}_{0}$, $a_0 = 1$ and $c_0 = d_0 = e_0 = 0$. This is a special case of a multivariate Tutte
polynomial \cite{Sokal05}. If $a_i = \cdots = e_i \equiv p$ ($0 \leq i \leq n$), we recover Chang and
Shrock's result (appendix 4.2 of \cite{ChangShrockA}) $\widehat{{\mathcal G}}_A(z) = \widehat{{\mathcal
N}}_A(z)/\widehat{{\mathcal D}}_A(z)$ with
\begin{eqnarray} \widehat{{\mathcal N}}_A(z) & = & p + p^3 \,
(1-p) \, (4 - 3 \, p)
\,z  ,\\
\widehat{{\mathcal D}}_A(z) & = & 1 - p^2\,\left( 12 - 26\,p + 21\,p^2 - 6\,p^3 \right) \,z \nonumber  \\
& &  + 2\, p^5\, {\left( 1 - p \right) }^3\,\left( 2 - p \right) \,z^2 .
\end{eqnarray}
The asymptotic power-law scaling factor is controlled by $\zeta_{+} = \frac{1}{2} \, p^2 \, ( 12 - 26 \, p
+21 \, p^2 -6 \, p^3 + \sqrt{{\mathcal B}})$ with ${\mathcal B} = 144 - 640\,p + 1236\,p^2 - 1308\,p^3 +
793\,p^4 - 260\,p^5 + 36\,p^6$.

\section{Conclusion and perspectives}
\label{Conclusions}

The two-terminal reliability of undirected networks may be expressed by a product of transfer matrices, in
which each edge and node reliability is exactly taken into account. This result is easily extended to the
all-terminal reliability with nonuniform links, as well as to directed graphs. We can now go beyond
series-parallel simplifications and look for new (wider) families of exactly solvable, meshed architectures
that may be useful for general reliability studies (as building blocks for more complex networks), for the
enumeration of self-avoiding walks on lattices, and for percolation with imperfect bonds {\em and} sites.
Since the true generating function is itself a rational fraction, Pad\'{e} approximants should provide
efficient upper or lower bounds for these studies. Moreover, individual reliabilities can be viewed as
average values of random variables. Having access to {\em each} edge or node allows the introduction of
disorder or correlations in calculations. The location of complex zeros of the two-terminal reliability
polynomials exhibits numerous structure transitions, with the possible occurrence of isolated points,
convergence to segments of the real axis, and also an expansion from the origin as $\rho$ goes to 0 which
obeys power-law behaviours with rational exponents which may differ strongly for seemingly not too dissimilar
graphs. All critical values of the node reliability are actually algebraic values. Finally, in a more applied
perspective, let us mention that the failure frequency $\nu$ of a given connection is another important
performance index of networks. If equipment $i$ with reliability $p_i$ has a failure rate $\lambda_i$, $\nu =
\sum_i \, \lambda_i \, p_i \, \partial {\rm Rel}_{2}/\partial p_i$. The matrix factorization makes the
calculation straightforward, since each $p_i$ appears in one transfer matrix only.

\ack

Very helpful information and suggestions by N.~L.~Biggs, N.~Bontemps, R.~Combescot, \'{E}.~Didelet, M.~Ducloy, F.~Glas, and G.~Patriarche are gratefully acknowledged.

\clearpage

\appendix

\section{A few recipes on the determination of the complex zeros of two-variate polynomials}

Our method relies on well-known results for the zeros of recursively defined one-parameter polynomials \cite{Biggs,ChangShrockA,Salas,Beraha78}. Since we are dealing here with two-variate ($p$,$\rho$) polynomials, let us outline the procedure used to obtain the figures, the critical values, and the asymptotic expansions given in the text.

\subsection{Determination of the two-variate polynomial}

As shown by many published studies, the convergence of the zeros to limiting sets of algebraic curves is already apparent for $n$ roughly equal to 50. To be on the safe side, we calculated these polynomials for $n = 100$ or $n = 150$ in order to (i) be very close to the asymptotic limit (ii) get a good sampling of the zeros, since --- especially in the small-$\rho$ limit --- they are not uniformly distributed over the asymptotic curves when $n \to \infty$ (see figures~\ref{ZerosStreet3NfnEQ0rho1.0}--\ref{ZerosStreet3NfnEQ0rho0.1} and \ref{ZerosStreet3NfnNEQ0rho0.01}).

We have calculated these polynomials using {\sc Mathematica} and recursion relations based on the denominator of the generating function. If
\begin{equation}
{\mathcal D}(z) = 1 + b_1(p,\rho) \, z + b_2(p,\rho) \, z^2 + \cdots + b_m(p,\rho) \, z^m \, ,
\label{denominateur D(z) general}
\end{equation}
then
\begin{equation}
{\rm Rel}_2^{(n)} = - b_1(p,\rho) \, {\rm Rel}_2^{(n-1)} - b_2(p,\rho) \, {\rm Rel}_2^{(n-2)} - \cdots - b_m(p,\rho) \, {\rm Rel}_2^{(n-m)} \, .
\label{recurrence Rel2}
\end{equation}
Knowledge of the first $m$ polynomials deduced from the generating function allows the quick determination of ${\rm Rel}_2^{(n)}$ for a given $\rho$. $\rho$ has not been kept as a parameter because of the explosion in the number of terms, but has been given rational values in order to prevent numerical errors; this gives polynomials with integral coefficients that may be very large (hundreds of digits sometimes). Their zeros have been obtained using {\sc Mathematica}'s routine \verb?NSolve?, the accuracy of which must be set accordingly (higher than hundreds of digits).

\subsection{Limiting curves and isolated zeros}

The zeros of recursively defined (one-parameter) polynomials mostly tend to aggregate close to curves such as (at least) two eigenvalues have the same modulus (the largest one for all the eigenvalues). Assuming that the ratio of the two eigenvalues is equal to $e^{{\rm i} \, \theta}$, we can write
\begin{equation}
{\mathcal D}(z) = (1 - \zeta \, e^{{\rm i} \, \theta/2} \, z) \, (1 - \zeta \, e^{-{\rm i} \, \theta/2} \, z) \, (1 - \widetilde{b}_{1} \, z - \cdots - \widetilde{b}_{m-2} \, z^{m-2}),
\label{D(z) avec les deux racines de meme module}
\end{equation}
which must be compared with (\ref{denominateur D(z) general}). Elimination of $\zeta$ and the $\widetilde{b}_{k}$'s leads to a (polynomial) relationship between $p$, $\rho$ and even powers of $t = \cos(\theta/2)$. Replacing $t$ by the more practical $T = \cos \theta$ gives a polynomial constraint ${\mathcal C}(p, \rho, T) = 0$. However, the true limiting curves are defined by only a subset of this constraint's many solutions for a given $\rho$ and $T$ ranging from -1 to +1, because $|\zeta|$ must be the largest. In this context, it does no harm to investigate special points of these curves.

\subsubsection{Double roots of ${\mathcal D}(z) = 0$}

In our case studies the endpoints of the limiting curves are such that both roots are equal ($\theta = 0$ or equivalently $T = +1$): they are thus obtained from a subset of the solutions of ${\widehat {\mathcal C}}(p, \rho) = {\mathcal C}(p, \rho, T = +1) = 0$. For the $K_4$-ladder with $\rho = 1$, this leads to
\begin{eqnarray}
0 & = & (2 - p )^2 \, (1-p)^4 \, (1-p+p^2)^2 \, \nonumber \\
& & (4 + 32\,p^2 - 204\,p^3 + 452\,p^4 - 516\,p^5 + 329\,p^6 - 112\,p^7 + 16\,p^8) \,
\label{K4 rho egal 1 T egal 1}
\end{eqnarray}
which gives the true endpoints of figure~\ref{ZerosRel2K4rho1.0}: $-0.1175415 \pm {\rm i} \, 0.2041183$, $0.7609223 \pm {\rm i} \, 0.5877642$, $1.343654 \pm {\rm i} \, 0.3456238$, and $1.512965 \pm {\rm i} \, 0.4931547$ (all the solutions are actually roots of the polynomial of degree 8). A quicker way to find these endpoints is to investigate when ${\mathcal D}(z)$ and $\displaystyle \frac{\partial {\mathcal D}(z)}{\partial z}$ are both equal to zero. Elimination of $z$ from these two equations leads to the desired ${\widehat {\mathcal C}}(p, \rho)$, or more accurately, to a product of two-variate polynomials. Confrontation with numerical estimates of the zeros allows to remove spurious solutions.

\subsubsection{Opposite roots of ${\mathcal D}(z) = 0$}

While they are usually not associated to remarkable points in the numerical plots of the complex zeros, they are nonetheless quite useful. Indeed, they pinpoint the limiting curves and can be obtained more easily because they satisfy ${\mathcal D}(z) = 0$ and ${\mathcal D}(-z) = 0$. Considering the even and odd components of ${\mathcal D}(z)$ as functions  of $Z = z^2$ and performing the elimination of $Z$ gives a new constraint ${\widetilde {\mathcal C}}(p, \rho)$, which is nothing but ${\mathcal C}(p, \rho, T = -1) = 0$. This task is simpler because the degree of the polynomials has been divided by two   (this definitely helps because even computer-assisted computations become ugly when the degree of ${\mathcal D}(z)$ increases). A few real zeros may correspond to opposite roots. For the $K_4$-ladder with $\rho = 1$, -0.2430623 and 1.527648 are indeed two such examples of intersections of the curves with the real axis, which may ultimately be tracked down to solutions of $-2 - 4 \, p + 14 \, p^2 -13 \, p^3 + 4 \, p^4 = 0$ (see figure \ref{ZerosRel2K4rho1.0}).

\subsubsection{Real roots and segments on the real axis}

They are frequent features of the complex zeros' structure. We mentioned in the previous paragraph that algebraic curves may intersect the real axis at a given $p$, the location of which can be traced back to particular roots of ${\mathcal D}(z) = 0$. Whole segments of the real (positive or negative) axis may also occur for some graphs (see figures~\ref{ZerosStreet3NfnEQ0rho0.6} and \ref{ZerosStreet3NfnNEQ0rho0.3}). It happens when, for a fixed $\rho$, two complex conjugate eigenvalues have the largest modulus for an extended range of real $p$'s. The proper assessment of the endpoints of this segment generally requires careful, numerical tests of the roots of ${\mathcal D}(z) = 0$. The existence of segments of the real axis may be restricted to a limited range of $\rho$'s or may persist down to $\rho \to 0$; it depends on the graph under consideration. When an algebraic curve ({\em and its symmetrical twin with respect to the real axis)} crosses the real axis, we have ${\mathcal C}(p, \rho, T) = 0$ and $\displaystyle \frac{\partial {\mathcal C}}{\partial p}(p, \rho, T) = 0$, because $p$ is a double (real) root at the intersection. The elimination of $T$ gives another polynomial constraint between $p$ and $\rho$.

\subsubsection{Isolated zeros, intersections with the imaginary axis, and roots of higher order}

Isolated zeros correspond to values of $p$ and $\rho$ such that the residue of the generating function --- taken at one of the eigenvalues of largest modulus --- simply vanishes. This implies that ${\mathcal D}(z)$ and ${\mathcal N}(z)$ are both equal to zero. Here again, the elimination of $z$ gives a constraint between $p$ and $\rho$. In the $K_4$-ladder and the $K_3$-cylinder with $f_i \neq 0$, this leads to (\ref{equation points isoles}) and (\ref{equation points isoles K3 fi neq 0}), respectively. 

In a few cases (see figures~\ref{ZerosStreet3NfnEQ0rho1.0} and \ref{ZerosStreet3NfnNEQ0rho1.0}), algebraic curves intersect the imaginary axis, even for vanishing $\rho$. Noting that if $p$ is a solution, then so is $-p$, we get a new constraint allowing the elimination of $T$.

In yet other instances, sets of algebraic curves join at triple points (see figures~\ref{ZerosStreet3NfnEQ0rho1.0}--\ref{ZerosStreet3NfnEQ0rho0.1}). This occurs when three roots of ${\mathcal D}(z) = 0$ have the same modulus. These points are usually harder to pinpoint in practice, especially away from the real axis.

\subsection{Critical values}

Changes --- sometimes quite drastic --- in the global structure of the complex zeros occur at particular values of $\rho$: the apparition or disappearance of real segments, isolated zeros, and small closed curves. These changes take place when, as $\rho$ varies, different pairs of eigenvalues have the largest modulus. Such a situation may be described in the following, simplified way. Let us assume that a particular point of the complex zeros' structure is described by ${\mathcal C}_1(p, \rho) = 0$. As $\rho$ decreases, this feature's origin changes and can be traced back to {\em another} constraint   ${\mathcal C}_2(p, \rho) = 0$. At the critical ($p_c,\rho_c$), both constraints must be satisfied. Elimination of one variable among $p$ and $\rho$ leads to the desired critical value. Since $\rho$ is kept real, we usually eliminate $p$. Not surprisingly, $\rho_c$ is a root of an algebraic equation, the degree and (integral) coefficients of which may become quite large. For instance, let us consider the apparition of the {\em third} isolated zero in the $K_4$-ladder configuration. Its existence is based on (\ref{equation points isoles}), which is apparently satisfied for $\rho \leq \rho_c$. When $\rho$ is slightly smaller than $\rho_c$, the isolated zero --- which remains on the real axis --- approaches the leftmost algebraic curve, which intersects the real axis at a point such that ${\mathcal C}(p, \rho, T = -1) = 0$. $\rho_c$ and the associated $p_c$ are therefore defined by their obeying the following two conditions, (\ref{equation points isoles}) and
\begin{eqnarray}
0 & = & 4 - 14\,p + 8\,p^2 \nonumber \\
& & + \left( 8\,p - 46\,p^2 + 130\,p^3 - 153\,p^4 + 80\,p^5 - 16\,p^6 \right) \,
   \rho \nonumber \\
& & + \left( 4\,p^2 + 18\,p^3 - 130\,p^4 + 249\,p^5 - 232\,p^6 \right. \nonumber \\
& & \left. \hskip5mm + 119\,p^7 - 33\,p^8 + 4\,p^9 \right) \,{\rho }^2 \, .
\label{equation T egal -1 K4-ladder}
\end{eqnarray}
The elimination of either $p$ or $\rho$ leads to the defining algebraic equation for the remaining parameter, which can be expressed as a product of polynomials. Comparison with the numerical data (one can always bracket $\rho_c$ or $p_c$ by trial and error) allows to select the relevant polynomial, given in (\ref{equation rho critique K4 troisieme zero isole}) and (\ref{equation p critique K4 troisieme zero isole}).

Obviously, the elimination procedure, which heavily relies on computer software ({\sc Mathematica} in the present case), works best when the degrees (in the variables to be eliminated) of the polynomials are not too large. A point may be worth mentioning: finding critical values involving only real $\rho_c$ and $p_c$ is usually much easier than for a real $\rho_c$ and complex conjugate $p_c$'s, because $p_c$ is associated with $T_c$ which is seldom equal to $\pm 1$. We have been able to calculate the critical $\rho_c$ corresponding to the apparition of the first two isolated (complex conjugate) zeros for the $K_4$-ladder, by considering the conditions ${\mathcal C}(p, \rho, T) = 0$ and (\ref{equation points isoles}), which can be decomposed in real and imaginary parts. This gives four equations and four parameters, namely $\rho$, $T$, ${\rm Re}(p)$, and ${\rm Im}(p)$. While it does not present any conceptual difficulty, this task may become numerically challenging since after each elimination procedure, the degrees in the remaining variables have a tendency to `explode'. Suffice it to say that the polynomial defining this critical $\rho_c$ is of degree 65, much larger than the degree 10 exhibited by (\ref{equation rho critique K4 troisieme zero isole}--\ref{equation p critique K4 troisieme zero isole}).

\subsection{Asymptotic expansions}

Our general method is to first assess numerically the expansion rate of the different substructures, which must behave as a negative fractional power of $\rho$ (because of the polynomial constraints in $p$ and $\rho$). This can be done by calculating the complex zeros for $\rho$ equal to $10^{-3}$, $10^{-6}$, etc. For instance, we infer from numerical calculations that the isolated zeros move from the origin with an expansion rate proportional to $\rho^{-1/3}$. Setting $p = \chi \, \rho^{-1/3}$ in (\ref{equation points isoles}) gives to lowest order $0 = 2 + 4 \, \chi^3 + {\mathcal O}(\rho^{1/3})$ and implies that $\displaystyle \chi^3 = -\frac{1}{2}$. The leading term is therefore easily obtained, down to its prefactor (note the symmetry of order 3 lying at the heart of the triplet of isolated zeros). The following terms of the asymptotic expansion may be deduced iteratively in a straightforward way.

As regards the sets of algebraic curves, the procedure is identical, with possibly different exponents. The `best' equation to start with is obtained for opposite roots (see above). For instance, setting $p = \chi \, \rho^{-1/4}$ in (\ref{equation T egal -1 K4-ladder}) gives $\displaystyle 0 = 8 \, \chi^2 \, \frac{1 - 2 \, \chi^4}{\sqrt{\rho}} + {\mathcal O}(\rho^{-1/4})$, implying $\displaystyle \chi^4 = \frac{1}{2}$.

Note that because the asymptotic structure is not strictly circular, the following terms of the expansion may depend on the argument (not only on the modulus) of the leading term of $\chi$. Finally, in such cases as the $K_3$-cylinder with $f_i = 0$, the above procedure gives several possible analytical solutions for $\chi$ with an expansion rate in $\rho^{-1/2}$, with very close numerical values which makes the correct identification of the true prefactor quite tedious. After careful numerical tests, we finally identified the expansion rate as $\displaystyle \frac{\left(5 - \sqrt{17} \right)^{1/4}}{\sqrt{2 \, \rho}}$.

\section{Transfer matrix for the $K_3$-cylinder}

The elements $m_{k,l}$ of the $13 \times 13$ transfer matrix $\widetilde{M}_i$ are
\begin{eqnarray*}
m_{1,1} & = & a_i \, S_i \\
m_{1,2} & = & c_i \, S_i \, T_i \, (b_i + d_i \, f_i \, U_i - b_i \, d_i \, f_i \, U_i) = - m_{4,2} \\
m_{1,3} & = & e_i \, S_i \, U_i \, (f_i + b_i \, d_i \, T_i - b_i \, d_i \, f_i \, T_i) \\
& & = - m_{5,3} = - m_{11,3} \\
m_{1,4} & = & a_i \, m_{1,2} =  m_{2,4} \\
m_{1,5} & = & a_i \, m_{1,3} = m_{3,11} = - m_{5,11} = - m_{11,5} \\
m_{1,6} & = & a_i \, c_i \, e_i \, S_i \, T_i \, U_i \, (d_i + b_i \, f_i - b_i \, d_i \, f_i) = - m_{11,6} \\
m_{1,7} & = & c_i \, e_i \, S_i \, T_i \, U_i \, (b_i \, d_i + b_i \, f_i + d_i \, f_i - 2 \, b_i \, d_i \, f_i) \\
& & = m_{13,7} = - m_{4,7} = - m_{11,7} \\
m_{1,8} & = & a_i \, m_{1,7} = m_{2,6} = m_{2,8} = m_{3,13} = m_{8,13} \\
& & = - m_{5,13} = - m_{7,13} = - m_{9,6} = - m_{9,8} = - m_{11,8} \\
m_{2,1} & = & a_i \, S_i \, T_i \, (b_i + d_i \, f_i \, U_i - b_i \, d_i \, f_i \, U_i) = - m_{4,1} \\
m_{2,2} & = & c_i \, T_i \\
m_{2,3} & = & e_i \, T_i \, U_i \, (d_i + b_i \, f_i \, S_i - b_i \, d_i \, f_i \, S_i) \\
& & = - m_{7,3} = - m_{9,3} \\
m_{2,5} & = & a_i \, e_i \, S_i \, T_i \, U_i \, (b_i \, d_i + b_i \, f_i + d_i \, f_i - 2 \, b_i \, d_i \, f_i) \\
& & = - m_{9,5} \\
m_{2,9} & = & c_i \, m_{2,3} = m_{3,9} \\
m_{2,10} & = & a_i \, (1 - b_i) \, c_i \, (1-d_i) \, e_i \, f_i \, S_i \, T_i \, U_i  \\
& & = m_{8,10} = m_{9,12} = m_{13,10} = - m_{4,10} \\
& & = - m_{7,10} = - m_{9,10} \\
m_{3,1} & = & a_i \, S_i \, U_i \, (f_i + b_i \, d_i \, T_i - b_i \, d_i \, f_i \, T_i) \\
& & = - m_{5,1} = - m_{11,1} \\
m_{3,2} & = & c_i \, T_i \, U_i \, (d_i + b_i \, f_i \, S_i - b_i \, d_i \, f_i \, S_i) \\
& & = - m_{7,2} = - m_{9,2} \\
m_{3,3} & = & e_i \, U_i \\
m_{3,4} & = & a_i \, c_i \, S_i \, T_i \, U_i \, (b_i \, d_i + b_i \, f_i + d_i \, f_i - 2 \, b_i \, d_i \, f_i) \\
m_{3,12} & = & a_i \, c_i \, e_i \, S_i \, T_i \, U_i \, (b_i + d_i \, f_i - b_i \, d_i \, f_i) \\
& & = m_{8,12} = - m_{5,12} = - m_{7,12} \\
m_{4,3} & = & - e_i \, S_i \, T_i \, U_i \, (b_i \, d_i + b_i \, f_i + d_i \, f_i - 2 \, b_i \, d_i \, f_i) \\
& & = - m_{13,3} \\
m_{4,4} & = & a_i \, c_i \, S_i \, T_i \, (1 - 2 \, b_i - 2 \, d_i \, f_i \, U_i + 2 \, b_i \, d_i \, f_i \, U_i ) \\
m_{4,5} & = & a_i \, e_i \, S_i \, T_i \, U_i \, (d_i - 2 \, b_i \, d_i - b_i \, f_i - 2 \, d_i \, f_i \\
& & + 3 \, b_i \, d_i \, f_i) = - m_{13,5} \\
m_{4,6} & = & a_i \, c_i \, e_i \, S_i \, T_i \, U_i \, (- b_i \, d_i + f_i - 2 \, b_i \, f_i - 2 \, d_i \, f_i \\
& & + 3 \, b_i \, d_i \, f_i) = - m_{13,6} \\
m_{4,8} & = & a_i \, c_i \, e_i \, S_i \, T_i \, U_i \, (d_i - 2 \, b_i \, d_i + f_i - 2 \, b_i \, f_i \\
& & - 3 \, d_i \, f_i + 4 \, b_i \, d_i \, f_i) = - m_{13,8} \\
m_{4,9} & = & (1 - b_i) \, c_i \, (1-d_i) \, e_i \, f_i \, S_i \, T_i \, U_i = m_{12,9} \\
m_{5,2} & = & - c_i \, S_i \, T_i \, U_i \, (b_i \, d_i + b_i \, f_i + d_i \, f_i - 2 \, b_i \, d_i \, f_i) \\
& & = m_{11,2} \\
m_{5,4} & = & a_i \, c_i \, S_i \, T_i \, U_i \, (d_i - 2 \, b_i \, d_i - b_i \, f_i - 2 \, d_i \, f_i \\
& & + 3 \, b_i \, d_i \, f_i) = m_{11,4} \\
m_{5,5} & = & a_i \, e_i \, (1 - f_i) \, S_i \, U_i \, (1 - b_i \, d_i \, T_i) = m_{11,11} \\
m_{5,6} & = & a_i \, b_i \, c_i \, (1-d_i) \, e_i \, (1 - f_i) \, S_i \, T_i \, U_i \\
m_{5,7} & = & c_i \, (b_i + d_i - 2 \, b_i \, d_i) \, e_i \, (1 - f_i) \, S_i \, T_i \, U_i \\
m_{5,8} & = & a_i \, m_{5,7} = m_{10,6} = m_{10,8} = m_{11,13} \\
m_{5,9} & = & - c_i \, e_i \, S_i \, T_i \, U_i \, (d_i + b_i \, f_i - b_i \, d_i \, f_i) \\
m_{6,2} & = & (1 - b_i) \, c_i \, (1-d_i) \, f_i \, S_i \, T_i \, U_i = m_{12,2} \\
m_{6,3} & = & b_i \, (1 - d_i) \, e_i \, (1-f_i) \, S_i \, T_i \, U_i = m_{10,3} \\
m_{6,4} & = & a_i \, m_{6,2} \\
m_{6,5} & = & a_i \, m_{6,3} = m_{9,11} \\
m_{6,6} & = & a_i \, c_i \, (1 - d_i) \, e_i \, (1 - b_i \, f_i) \, S_i \, T_i \, U_i \\
m_{6,7} & = & c_i \, (1 - d_i) \, e_i \, (b_i + f_i - 2 \, b_i \, f_i) \, S_i \, T_i \, U_i \\
m_{6,8} & = & a_i \, m_{6,7} = m_{7,6} = m_{7,8} = m_{9,13} \\
m_{7,1} & = & - a_i \, S_i \, T_i \, U_i \, (b_i \, d_i + b_i \, f_i + d_i \, f_i - 2 \, b_i \, d_i \, f_i) \\
& & = m_{9,1} = - m_{8,1} \\
m_{7,4} & = & a_i \, c_i \, S_i \, T_i \, U_i \, (- b_i \, d_i + f_i - 2 \, b_i \, f_i - 2 \, d_i \, f_i \\
& & + 3 \, b_i \, d_i \, f_i) = m_{9,4} \\
m_{7,5} & = & a_i \, (1 - d_i) \, e_i \, (b_i + f_i - 2 \, b_i \, f_i) \, S_i \, T_i \, U_i \\
m_{7,9} & = & c_i \, e_i \, T_i \, U_i \, (1 - 2 \, d_i - 2 \, b_i \, f_i \, S_i + 2 \, b_i \, d_i \, f_i \, S_i ) \\
& & = m_{9,9}\\
m_{7,11} & = & - a_i \, e_i \, S_i \, T_i \, U_i \, ( b_i \, d_i + f_i - b_i \, d_i \, f_i) = - m_{8,11} \\
m_{8,2} & = & - c_i \, S_i \, T_i \, U_i \, (- b_i \, d_i + f_i - 2 \, b_i \, f_i - 2 \, d_i \, f_i \\
& & + 3 \, b_i \, d_i \, f_i) = m_{13,2} \\
m_{8,3} & = & - e_i \, S_i \, T_i \, U_i \, (b_i - 2 \, b_i \, d_i - 2 \, b_i \, f_i - d_i \, f_i \\
& & + 3 \, b_i \, d_i \, f_i) = m_{13,2} \\
m_{8,4} & = & - a_i \, c_i \, S_i \, T_i \, U_i \, (d_i - 2 \, b_i \, d_i + 2 \, f_i - 3 \, b_i \, f_i \\
& & - 4 \, d_i \, f_i + 5 \, b_i \, d_i \, f_i) \\
m_{8,5} & = & - a_i \, e_i \, S_i \, T_i \, U_i \, (2 \, b_i + d_i - 3 \, b_i \, d_i + f_i - 3 \, b_i \, f_i \\
& & - 2 \, d_i \, f_i + 4 \, b_i \, d_i \, f_i) \\
m_{8,6} & = & - 2 \, a_i \, m_{6,7} \\
m_{8,7} & = & - c_i \, e_i \, S_i \, T_i \, U_i \, (2 \, b_i + d_i - 3 \, b_i \, d_i + f_i - 3 \, b_i \, f_i \\
& & - 2 \, d_i \, f_i + 4 \, b_i \, d_i \, f_i) \\
m_{8,8} & = & - a_i \, c_i \, e_i \, S_i \, T_i \, U_i \, (-1 + 3 \, b_i + 2 \, d_i - 4 \, b_i \, d_i + 3 \, f_i \\
& & - 5 \, b_i \, f_i - 4 \, d_i \, f_i + 6 \, b_i \, d_i \, f_i) \\
m_{8,9} & = & - c_i \, e_i \, S_i \, T_i \, U_i \, (- d_i + f_i - 2 \, b_i \, f_i - d_i \, f_i \\
& & + 2 \, b_i \, d_i \, f_i) \\
m_{10,1} & = & a_i \, (1 - b_i) \, d_i \, (1-f_i) \, S_i \, T_i \, U_i = m_{12,1} \\
m_{10,4} & = & c_i \, m_{10,1} \\
m_{10,5} & = & a_i \, (b_i + d_i - 2 \, b_i \, d_i) \, e_i  \, (1 - f_i) \, S_i \, T_i \, U_i = - m_{13,11} \\
m_{10,9} & = & b_i \, c_i \, (1 - d_i) \, e_i \, (1-f_i) \, S_i \, T_i \, U_i = m_{11,9} \\
m_{10,10} & = & a_i \, (1 - b_i) \, c_i \, (1 - d_i) \, e_i \, (1-f_i) \, S_i \, T_i \, U_i \\
m_{11,12} & = & a_i \, (1 - b_i) \, c_i \, d_i \, e_i \, (1-f_i) \, S_i \, T_i \, U_i \\
m_{12,4} & = & a_i \, (1 - b_i) \, c_i \, (d_i + f_i - 2 \, d_i \, f_i) \, S_i \, T_i \, U_i \\
m_{12,11} & = & e_i \, m_{10,1} \\
m_{12,12} & = & a_i \, (1 - b_i) \, c_i \, e_i \, (1 - d_i \, f_i) \, S_i \, T_i \, U_i \\
m_{12,13} & = & e_i \, m_{12,4} = - m_{13,12} \\
m_{13,1} & = & - a_i \, S_i \, T_i \, U_i \, (d_i - 2 \, b_i \, d_i - b_i \, f_i - 2 \, d_i \, f_i \\
& & + 3 \, b_i \, d_i \, f_i) \\
m_{13,4} & = & - a_i \, c_i \, S_i \, T_i \, U_i \, (2 \, d_i - 3 \, b_i \, d_i + 2 \, f_i - 3 \, b_i \, f_i \\
& & - 5 \, d_i \, f_i + 6 \, b_i \, d_i \, f_i) \\
m_{13,9} & = & c_i \, (1 - d_i) \, e_i \, (- b_i - 2 \, f_i + 3 b_i \, f_i) \, S_i \, T_i \, U_i \\
m_{13,13} & = & - a_i \, c_i \, e_i \, S_i \, T_i \, U_i \, (-1 + 2 \, b_i + 2 \, d_i - 3 \, b_i \, d_i + 2
\, f_i \\
& & - 3 \, b_i \, f_i - 3 \, d_i \, f_i + 4 \, b_i \, d_i \, f_i)
\end{eqnarray*}

All the following matrix elements are equal to zero: $m_{1,9}$, $m_{1,10}$, $m_{1,11}$, $m_{1,12}$,
$m_{1,13}$, $m_{2,7}$, $m_{2,11}$, $m_{2,12}$, $m_{2,13}$, $m_{3,5}$, $m_{3,6}$, $m_{3,7}$, $m_{3,8}$,
$m_{3,10}$, $m_{4,11}$, $m_{4,12}$, $m_{4,13}$, $m_{5,10}$, $m_{6,1}$, $m_{6,9}$, $m_{6,10}$, $m_{6,11}$,
$m_{6,12}$, $m_{6,13}$, $m_{7,7}$, $m_{9,7}$, $m_{10,2}$, $m_{10,7}$, $m_{10,11}$, $m_{10,12}$, $m_{10,13}$,
$m_{11,10}$ $m_{12,3}$, $m_{12,5}$, $m_{12,6}$, $m_{12,7}$, $m_{12,8}$, $m_{12,10}$.


Note that for $i = 0$, one must set $a_0 =1$ and $c_0 = e_0 = 0$.

\section*{References}

\end{document}